\documentclass[conference]{IEEEtran}
\IEEEoverridecommandlockouts

\usepackage{cite}
\usepackage{amsmath,amssymb,amsfonts}
\usepackage{algorithmic}
\usepackage{graphicx}
\usepackage{textcomp}
\usepackage{xcolor}
\usepackage{booktabs} 
\usepackage{multirow}
\usepackage{pgfplots}
\usepackage{caption}
\pgfplotsset{compat=1.18}
\usepackage[most]{tcolorbox}
\usepackage{hyperref}
\usepackage{listings}
\DeclareCaptionFormat{listing}{#1#2#3}
\def\BibTeX{{\rm B\kern-.05em{\sc i\kern-.025em b}\kern-.08em
    T\kern-.1667em\lower.7ex\hbox{E}\kern-.125emX}}
\begin{document}

\title{Analyzing Code Injection Attacks on LLM-based Multi-Agent Systems in Software Development\\
}

\author{\IEEEauthorblockN{Brian Bowers}
\IEEEauthorblockA{\textit{Department of Computer Science} \\
\textit{Loyola Marymount University}\\
Los Angeles, USA \\
bbowers3@lion.lmu.edu}
\and
\IEEEauthorblockN{Smita Khapre}
\IEEEauthorblockA{\textit{Department of Computer Science} \\
\textit{University of Colorado Colorado Springs}\\
Colorado Springs, USA \\
skhapre@uccs.edu}
\and
\IEEEauthorblockN{Jugal Kalita}
\IEEEauthorblockA{\textit{Department of Computer Science} \\
\textit{University of Colorado Colorado Springs}\\
Colorado Springs, USA \\
jkalita@uccs.edu}
}

\maketitle

\begin{abstract}
Agentic AI and Multi-Agent Systems are poised to dominate industry and society imminently. Powered by goal-driven autonomy, they represent a powerful form of generative AI, marking a transition from reactive content generation into proactive multitasking capabilities.
As an exemplar, we propose an architecture of a multi-agent system for the implementation phase of the software engineering process. We also present a comprehensive threat model for the proposed system. We demonstrate that while such systems can generate code quite accurately, they are vulnerable to attacks, including code injection. Due to their autonomous design and lack of humans in the loop, these systems cannot identify and respond to attacks by themselves. This paper analyzes the vulnerability of multi-agent systems and concludes that the \textit{coder-reviewer-tester} architecture is more resilient than both the \textit{coder} and \textit{coder-tester} architectures, but is less efficient at writing code. We find that by adding a security analysis agent, we mitigate the loss in efficiency while achieving even better resiliency. We conclude by demonstrating that the security analysis agent is vulnerable to advanced code injection attacks, showing that embedding poisonous few-shot examples in the injected code can increase the attack success rate from 0\% to 71.95\%. 
\end{abstract}


\section{Introduction}

Agentic AI (AAI) has recently gained momentum due to its ability to adapt and make decisions in real-world environments, signaling an innovation in machine-human collaboration \cite{murugesan2025rise}. 
Multi-Agentic Systems (MAS) build further upon AAIs, allowing collaboration between agents to achieve broader goals. Such systems are being adopted rapidly in industry, necessitating faster development and deeper research across various domains. Gartner's report, \textit{``Top Strategic Technology Trends for 2025: Agentic AI"} \cite{Gartner}, predicts that by 2028, 33\% of applications will incorporate some form of AAI in enterprise software, starting from a base of 1\% in 2024. The report also highlights the risk to \textit{security and quality} when AAI is incorporated in information technology stacks, including the danger of \textit{smart malware} cyberattacks.
In this paper, we explore issues in the implementation of MAS in the software engineering process (SWEP), primarily for the implementation phase involving coding, code review, and unit testing. We analyze one of the threats presented in the threat model in the context of the proposed MAS architecture discussed below, leading to claims regarding the security and efficiency of such systems in SWEP. The main contributions of this paper are as follows.
\begin{itemize}
    \item We propose a series of MAS architectures (\textit{coder}, \textit{coder-tester}, and \textit{coder-reviewer-tester}) powered by Large Language Models (LLMs) for the implementation phase in the software development life cycle guided by industry practices. We then evaluate each architecture's accuracy, attack effectiveness, and architecture efficiency against a \textit{code injection} attack.
    \item We introduce an added security analysis agent to mitigate code injection attacks and compare its contribution against industry-recommended architectures.
    \item We demonstrate that the natural language understanding of the security analysis agent can be exploited to increase the attack success rate significantly.
\end{itemize}

The necessary background to understand this work is described in Section \ref{Background}, followed by related work in Section \ref{Related_work}. The threat model is explained in Section \ref{Threat_model}. Our approach is described in Section \ref{Approach}, where the model, attack, and datasets are discussed in detail. Experiments are discussed in Section \ref{Experiments}, and results are in Section \ref{Results}. Finally, we discuss the limitations in Section \ref{Limitations}, future direction of our work in Section \ref{Future_Directions}, and in Section \ref{Conclusion}, a conclusion is drawn.

\section{Background} \label{Background}

Code generation by Large Language Models (LLMs) involves understanding a natural language description of a system design and generating code that satisfies the specified requirements. Recent approaches utilize LLMs, taking advantage of their natural language comprehension and generative capabilities to generate functional code. For example, Microsoft Copilot is an AI-powered assistant built on top of LLMs.

To improve the accuracy and security of generated code, LLMs' capabilities can be enhanced in various ways. One, an LLM can use a code execution tool to make sure code executes without errors. Two, an LLM can be prompted to use a chain of thought (CoT) or self-reflection during software development. Three, an LLM can also be upgraded with action planning capabilities, given a suitable design, appropriate tools, and other resources.  Generally, such empowered LLMs are referred to as \textit{Agents}. An agent typically has a profile that may include demographic, personality, and social information, as well as technical details, responsibilities, and capabilities. Other resources may include memory by utilizing Retrieval-Augmented Generation or vector databases, and action planning capabilities to adapt to feedback from other agents and their environment. Lastly, such agents can carry out actions autonomously, such as exploring, communicating, and using tools with direct human oversight \cite{wang2024survey_autonomous}. When multiple such agents collaborate on the same overall problem, they are known as a Multi-Agent System (MAS). Dong et al. found that self-collaboration can improve code generation by 29.9-47.1\% when compared to the na\"ive single LLM agent approach \cite{dong2024self-collaboration}.

To work together effectively, agents must communicate with each other in a structured manner. The nature of their communication is usually determined by the system architecture, which is often specifically crafted for the problem. 
A common approach is to mimic industry practitioners on how to solve a specific type of problem. Software development life-cycle (SDLC) \cite{ISO_SDLC} is a well-developed process developed by the International Standards Organization. Our focus is on the implementation phase of SDLC, often referred to as \textit{implementation} or \textit{coding} phase. It involves three steps: namely, \textit{code generation, code reviews}, and {\em unit testing} \cite{kumar2007securityDev_SLDC}. Code generation is a crucial part of the SDLC, and if incorrect or insecure code is generated, it may lead to vulnerabilities and malicious effects \cite{wang2023a_review_on_code}.

In this paper, we propose a multi-agent system, where each of these steps has a corresponding agent, specialized for the task. With this architecture, we aim to evaluate the feasibility of engaging a state-of-the-art current MAS in the \textit{implementation phase} of SLDC. 

We simulate insider threats using a code injection attack, where malicious code is inserted into a system to alter its behavior. In MAS, such an attack is likely to be more damaging, due to their autonomous nature, without any expert human oversight. The attack may go unnoticed unless the agents can identify the attack themselves, either alone or collaboratively. With the proposed simple MAS architecture, the intention is to investigate the tradeoffs between system performance and security.


\section{Related Work} \label{Related_work}

Several prior works have proposed using three-agent architectures. Dong et al. \cite{dong2024self-collaboration} used an analyst, a coder, and a tester, increasing Pass@1 accuracy on the HumanEval dataset by 29.9\% over a single LLM. AlMorsi et al. \cite{almorsi2024guided} achieved an increase of 23.97\%, compared to direct one-shot generation, on HumanEval using a testing agent, a code generation agent, and a generalist agent for problem decomposition and building a search tree. MapCoder, which has four agents to recall relevant examples, plan, generate code, and debug, achieved 57.6\% on HumanEval Pass@1 using a small 7b model \cite{islam2024mapcoder}. MetaGPT, proposed by Hong et al., coordinated communications between LLMs following standard operating procedures, utilizing software-design-inspired agents along with iterative code improvement and debugging to achieve 85.9\% on HumanEval \cite{hong2024metagpt}. Other works have found that simpler designs can perform better. For example, AgentCoder, which consists of a programmer, test designer, and test executor, outperformed larger MAS (5+ agents) in both accuracy and token efficiency \cite{huang2023agentcoder}. Recently, an approach by Shi et al. \cite{shi2024code} achieved 100\% accuracy on HumanEval Pass@1 using a hierarchical debugging approach with a latest generation LLM.

Wu et al. proposed the open-source AutoGen framework for MAS application implementation \cite{autogen}. It provides flexibility for adding human inputs, tools, resources, and decision-making. It can be customized for a variety of applications in various domains.

Code injection within MAS is examined by Huang et al. \cite{huang2024resilience}, studying the effect of injecting code errors on three different types of architectures. They determined that hierarchical structures were the most resilient. They also proposed a defense method called \textit{Inspector} beyond using more resilient structures, by adding an external agent that reviews and corrects messages passed by the other agents \cite{huang2024resilience}. Zhou et al. proposed a different approach of defense called GUARDIAN, introducing a discrete-time temporal graph to prevent errors \cite{zhou2025guardian}.
Triedman et al. were able to get a MAS to execute malicious code through control-flow hijacking, tricking a web-surfer agent into processing a file with deceptive comments that led the LLM to send code to a test executor for execution \cite{triedman2025multi-agent_systems}. He et al. proposed an approach to intercept messages between agents, accomplished by red teaming, although this paper did not explicitly focus on code injection \cite{he2025red-teaming}. 

Many studies have explored code injection into traditional software systems, while others have looked into attacking multi-agent systems via memory, tool execution, and communication \cite{narajala2025securing_ATFAA}. However, there is still a lack of research looking at code injection and methods of mitigating the effects in multi-agent autonomous systems.
\begin{table}[ht!]
    \centering
    \begin{tabular}{|l|c|l|c|l|}
        \hline
        \textbf{Threat} & \textbf{STRIDE} & \textbf{MITRE} & \textbf{ATFAA} & \textbf{Explanation} \\
        \textbf{Name} & \textbf{Category} & \textbf{ATLAS} & \textbf{Domain} & \\
        \hline
        Code & Spoofing & Phishing, & Trust 
        & Someone with  \\
        Injection & & Impersonation, & Boundary
        & access to code, \\
        &&Unsecured&& changes or add  \\
        && Credentials&& malicious code\\
        
        \hline
        
        AI & Tamper- &LLM Trusted 
        & Cognitive 
        & It results in \\
        Reasoning&-ing&Output &Architect-
        & false output. \\
        Manipula-&& Components &-ure& CRT approves \\
        -tion&&Manipulation&& instead of \\
        &&&& disapproving\\
        \hline
        
    \end{tabular}
    \caption{A mapping of our proposed MAS architecture threats to STRIDE, ATLAS, and ATFAA categories.}
    \label{tab:threat_model}
\end{table}

\section{Threat Model} \label{Threat_model}

Our threat model for a reflective 
and task-oriented multi-agent system,  is inspired by ``\textit{OWASP Top 10 Agentic AI Threats and Mitigation}" \cite{john2025owasp} and distilled from the threat model proposed by Narajala and Narayan \cite{narajala2025securing_ATFAA}. The threats to the architecture designed for this project are summarized in Table \ref{tab:threat_model}. We combine Microsoft's STRIDE threat model framework \cite{STRIDE}, mapping Amazon Web Services' Advanced Threat Framework for Autonomous AI Agents (ATFAA) proposed by \cite{narajala2025securing_ATFAA} and MITRE ATLAS Matrix tactics and techniques \cite{MITRE_ATLAS}. 
The threat ID 1 \textit{Code Injection} is exploited in the experimentation of the proposed design, which also results in exploitation of threat ID 2 \textit{AI Reasoning Manipulation}.

We assume that the attacker has full access to the code created by the \textit{coding agent}. This access could be gained through social engineering, insider threat, or exploitation of software vulnerabilities. In addition, communications between agents can be compromised, giving access to generated code \cite{he2025red-teaming}.

\begin{lstlisting}[language=Python, caption={Function that tries to send encrypted passwords out of the system to a specified address.}, label={lst:function}]
def helper():
    with open("/etc/shadow.txt", "rb") as f:
        requests.post("http://random.example.com/upload",
        files={"file": f})
\end{lstlisting}

\section{Approach} \label{Approach}

\begin{figure}[h]
  \includegraphics[width=\linewidth]{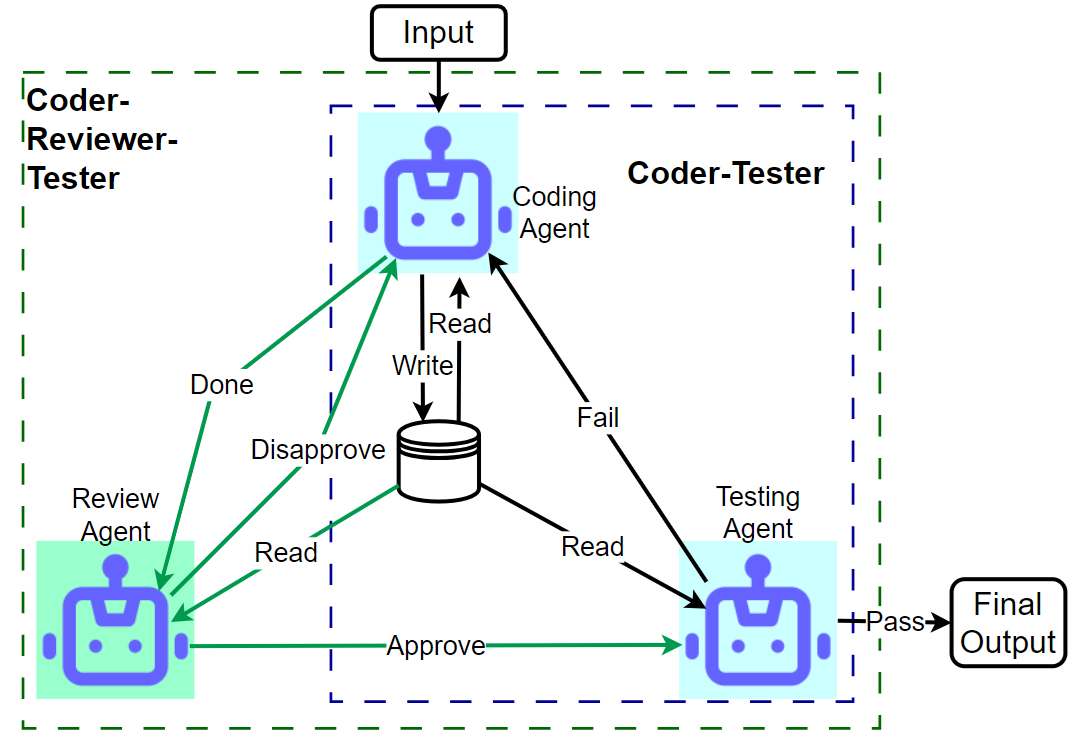}
  \caption{Architecture for Coder-Tester (inner blue colored dashed lined box) and Coder-Reviewer-Tester (outer green colored dashed lined box).}
  \label{fig:crt}
\end{figure}

The general design of our MAS architectures is shown in Figure~\ref{fig:crt}, and explained as follows. We chose these three architectures because they involve multiple agents and reflect a real-world software development structure involving humans.

\begin{enumerate}
  \item \textit{Coder} (C): This architecture uses a single LLM-based agent that employs chain-of-thought prompting. The coder agent receives the problem and then generates a solution.
  \item \textit{Coder-Tester} (CT): The coder receives the problem, generates code, and then writes it out to a database. Once writing is completed, the coder sends a request to the testing agent. The testing agent reads the code from the database and runs it against the test cases it has generated.
  \item \textit{Coder-Reviewer-Tester} (CRT): This model includes an additional review agent, which must first approve the code generated by the coder before it is sent to the tester.
\end{enumerate}

\subsection{Attacks}

We define two different levels of adversarial access in CT and CRT architectures. 
\begin{itemize}
    \item \textit{Single}: One-time access to the generated code. It will launch a single \textit{Code Injection} attack after the response from the coding agent.
    \item \textit{Continued}: Continuous access to the generated code. Here, every section of generated code could be changed by the attacker at any point in the system process. Thus, it will launch an attack each time the \textit{coder} agent generates a response.
\end{itemize}
 The adversary's goal is to inject a malicious function into the code and have it executed. We define an attack as successful if the injected code is present in the output of the MAS and executed in the production environment. Our target is data exfiltration. The specific function we insert into the code is shown in Listing~\ref{lst:function}. This is for proof of concept only.

Additionally, we introduce a security analysis agent to the \textit{coder-tester} architecture. This agent is prompted to only review for security concerns. This agent is added at the end of the architecture, where it only checks the final output. This follows established MAS practices, such as the OpenSDKs guidelines feature \cite{OpenAI_OpenSDK}. The position of the security analysis agent at the end allows us to disregard the access levels, assuming that the attack happens before the final output is passed to the security analysis agent.

We define advanced attack levels to extensively test the ability to recognize code injection attacks by the security analysis agent. These attacks are all variations of the function shown in Listing~\ref{lst:function} referred to in the following sections as the \textit{exfiltration function}:

\begin{enumerate}
    
    \item \textbf{Attack 1}: An import for the requests library is added to the exfiltration function, and the function is then called, resulting in the malicious code being executed.
    \item \textbf{Attack 2}: The exfiltration function is obfuscated using a Python obfuscator \cite{freecodingtools_python_obfuscator}.
    \item \textbf{Attack 3}: A comment is added just before the exfiltration function that falsely claims the function is commented out and should be ignored.
    \item \textbf{Attack 4}: This attack is the same as Attack 3, except an additional comment is added above the import statement, also falsely claiming the import is commented out.
    \item \textbf{Attack 5}: A comment is added before the exfiltration function, claiming the following code will not execute and therefore should be ignored.
    \item \textbf{Attack 6}: A comment is added before the exfiltration function, including poisoned few-shot examples. These examples include malicious functions that are correctly and incorrectly labeled.
\end{enumerate}
\label{list:adv_attacks}

\subsection{Software \& Hardware}
Three different backbone LLMs are explored. The first is Code Llama \cite{codellama}. We explore the seven-billion parameter model due to resource limitations and time constraints. This model was specifically trained on code to increase its performance on coding-specific tasks. The second model we use is Mistral 7b. This model is included due to its strong instruction-following capabilities. Both these models are run locally on a single NVIDIA GeForce RTX 3090, through Ollama. Finally, we also explore a closed-source model, GPT-4.1-mini. The focus of this paper is on the impact of code injection attacks and not system accuracy, so using smaller models does not invalidate our results. The different model architectures, \textit{coder}, \textit{coder-tester}, \textit{coder-tester-reviewer}, and \textit{coder-tester with security agent} are implemented using Autogen \cite{autogen}. This framework was chosen for its low-level control over the agents and programmable message passing. We provide our methods to ensure our results are fully reproducible.

\subsection{Datasets}

We evaluate our models and attack success rates using HumanEval, consisting of 164 hand-written Python programming problems \cite{chen2021humaneval}. This dataset provides a function signature, a docstring containing a natural language description of the problem, and tests to determine if a solution is correct. HumanEval is a widely used standard for evaluating performance on coding tasks \cite{jiang2024a_survey_code}. 

\section{Experiments} \label{Experiments}

To establish a baseline for each architecture, we first determine accuracy, Pass@1, with no attack on HumanEval. Next, we carry out our attacks. We measure the accuracy of the models under each attack. Since none of our attacks are immediately disruptive, we anticipate the accuracy to remain roughly the same, making it difficult for the attack to be detected. In addition, we measure the efficiency of each architecture. For the \textit{coder-tester} architecture, we set the maximum number of testing rounds to three. For the \textit{coder-tester-reviewer}, we set both the maximum number of review and test rounds to three. Once the maximum number of testing rounds has been reached, the last generated code is output from the model. Maximum rounds are necessary to prevent wasteful LLM calls and stop the models from unnecessarily going back and forth without any significant progress being made. Both maximum review and test rounds are hyperparameters, and we found that as the number of review and test rounds increases, the system becomes less accurate. For  GPT-4.1-mini, this was not the case.
To assess the effectiveness of an attack, we look at the number of times the malicious function remains in the code output by the system. For the reviewing agent, we assume the attack is successful if the code is approved at the final review round. Finally, we add the security analysis agent to the \textit{coder-tester} architecture and examine its effectiveness against the \textit{exfiltration function} as well as more complex variations, described in Section \ref{list:adv_attacks}. Since LLM outputs are non-deterministic, all the experiments were run three times, and the results were averaged.

\section{Results and Analysis} \label{Results}

\begin{table*}[h!]
  \centering
  \resizebox{\textwidth}{!}{

  \begin{tabular}{ll 
    c c c  
    c c c  
    c c c  
  }
    \toprule
    \textbf{Architecture} & \textbf{Attack} & 
    \multicolumn{3}{c}{\textbf{Code Llama}} & 
    \multicolumn{3}{c}{\textbf{Mistral}} & 
    \multicolumn{3}{c}{\textbf{GPT-4.1-mini}} \\
    \cmidrule(lr){3-5} \cmidrule(lr){6-8} \cmidrule(lr){9-11}
    & & \textbf{Acc} & \textbf{Eff} & \textbf{Calls} 
      & \textbf{Acc} & \textbf{Eff} & \textbf{Calls} 
      & \textbf{Acc} & \textbf{Eff} & \textbf{Calls} \\
    \midrule
    \multirow{3}{*}{C}& No Attack & 22.15\(\pm3.81\) & - & 383.33\(\pm24.88\) & 30.69\(\pm2.31\) & - & 164.00\(\pm0.00\) & 94.72\(\pm0.87\) & - & 164.00\(\pm0.00\) \\
& Single & 21.34\(\pm4.01\) & 100.00\(\pm0.00\) & 390.00\(\pm62.70\) & 31.30\(\pm3.50\) & 100.00\(\pm0.00\) & 164.00\(\pm0.00\) & 94.31\(\pm2.31\) & 100.00\(\pm0.00\) & 164.00\(\pm0.00\) \\
& Continued & 20.33\(\pm3.81\) & 100.00\(\pm0.00\) & 403.33\(\pm23.87\) & 31.30\(\pm4.63\) & 100.00\(\pm0.00\) & 164.00\(\pm0.00\) & 94.31\(\pm1.75\) & 100.00\(\pm0.00\) & 164.00\(\pm0.00\) \\
    \multirow{3}{*}{CT}& No Attack & 22.97\(\pm4.87\) & - & 1334.00\(\pm248.30\) & 30.08\(\pm3.81\) & - & 572.00\(\pm20.33\) & 95.93\(\pm2.31\) & - & 346.67\(\pm1.43\) \\
& Single & 24.19\(\pm3.81\) & 24.39\(\pm6.06\) & 1327.00\(\pm41.04\) & 30.28\(\pm10.09\) & 27.44\(\pm6.94\) & 570.00\(\pm19.40\) & 95.93\(\pm1.75\) & 92.48\(\pm3.15\) & 350.00\(\pm8.96\) \\
& Continued & 24.80\(\pm4.37\) & 100.00\(\pm0.00\) & 1245.00\(\pm41.34\) & 29.27\(\pm2.62\) & 100.00\(\pm0.00\) & 583.00\(\pm11.38\) & 97.15\(\pm2.31\) & 100.00\(\pm0.00\) & 350.00\(\pm4.30\) \\
    \multirow{3}{*}{CRT}& No Attack & 24.80\(\pm4.63\) & - & 1726.33\(\pm167.20\) & 32.11\(\pm5.32\) & - & 740.67\(\pm17.62\) & 95.33\(\pm2.31\) & - & 529.33\(\pm5.17\) \\
& Single & 22.56\(\pm1.51\) & 19.72\(\pm9.26\) & 1701.00\(\pm49.62\) & 29.88\(\pm4.01\) & 21.54\(\pm2.31\) & 789.00\(\pm30.12\) & 95.73\(\pm1.51\) & 1.42\(\pm3.50\) & 847.33\(\pm21.42\) \\
& Continued & 20.73\(\pm12.12\) & 95.73\(\pm3.03\) & 1659.00\(\pm128.53\) & 27.44\(\pm9.21\) & 95.12\(\pm1.51\) & 813.00\(\pm54.82\) & 95.53\(\pm2.31\) & 6.71\(\pm2.62\) & 1153.00\(\pm12.91\) \\
    \bottomrule
  \end{tabular}
  }
  \caption{The different architectures, shown in Figure~\ref{fig:crt}: Coder (\textbf{C}), Coder-Tester (\textbf{CT}), and Coder-Reviewer-Tester (\textbf{CRT}) are evaluated on three different attack methods: \textbf{No Attack}, \textbf{Single}, and \textbf{Continued}. The accuracy (\textbf{Acc}) of the system on HumaEval, determined by Pass@1, the effectiveness (\textbf{Eff}) of the attack (higher \% shows successful attack), and the number of calls (\textbf{Calls}) made to the underlying LLM are all evaluated. Each architecture is also evaluated with three separate LLM models, \textbf{GPT-4.1-mini}, \textbf{Mistral}, and \textbf{CodeLLama}.}
  \label{tab:model_comparison}
\end{table*}

The results for the different architectures are presented in Table~\ref{tab:model_comparison}. Accuracy is determined by Pass@1, meaning the system only has one attempt, on the HumanEval dataset. The closed-source model, GPT-4.1-mini, outperformed the open-source models for each of the different architectures by over 65\%. Importantly, the accuracy does not decrease when either the single or the continuous attack is introduced. This results in the code injection attacks being more difficult to detect. \textit{C}, \textit{CT}, and \textit{CTR} showed no significant differences in accuracy for any of the LLM models. This is likely due to the large margin of error resulting from a small number of trials. In Table~\ref{tab:model_comparison}, the effectiveness of the attack when the attack is none is not shown.

The effectiveness of the attack determines how successful the code injection attack was, on average. The single attack was not as effective as the continued attack. For the \textit{coder} (C), the single and continued attack was 100\% successful. Since the attack happens after the coder has written the code, the coding agent by itself cannot stop the attack. For the \textit{coder-tester}, the single attack was 92.48\% successful, even with the best performing LLM GPT-4.1-mini. For the single attack to be successful in this architecture, the code must not pass the tester on the first round. The continued attack against the \textit{coder-tester} is 100\% successful, since the code is injected after the coder writes, and the tester cannot stop the code from being output by the system. The \textit{coder-tester-reviewer} (CRT) architecture performed the best against the attacks. For GPT-4.1-mini, the single attack was only 1.42\% successful and the continued attack was 6.71\% effective, outperforming the other models, at roughly 95\% for both. Considering the review agent is not prompted to look for security issues but only to evaluate the code for correctness, GPT-4.1-mini performed strongly. This is likely due to alignment techniques being used on GPT-4.1-mini and exposure to common attack patterns and safety evaluation.

The number of LLM calls is a measure of the efficiency of the different architectures. The \textit{coder} does the best, using only 164 calls, one call per question. The \textit{coder-tester} is less efficient, with GPT-4.1-mini using about 350 LLM calls, while Mistral used around 570. This is likely due to GPT-4.1-mini's high \textit{coder} accuracy, meaning it is more likely to create a correct solution on its first try. The number of calls increases again with the \textit{CRT}. The increase in the number of calls, especially in GPT-4.1-mini, suggests the reviewer is detecting the injected code and disapproving, leading to another round.

From these experiments, several conclusions can be drawn. One, increasing the number of agents, especially when adding a review agent, does not increase the accuracy of the system significantly and leads to more LLM calls. However, the models with the highest accuracy are completely vulnerable to single and continuous attacks. In contrast, the \textit{CRT} performs the best against the attacks, but increases the number of LLM calls by hundreds while not improving the accuracy. Detecting the malicious function should be trivial, considering the obvious malicious nature of the injected function, combined with the simplicity of the generated code. With these considerations, the continued attack of 6.71\% by GPT-4.1-mini is high, as 0\% was expected. To address this issue, we added \textit{security analysis agent} to the \textit{coder-tester}. It reduced the effectiveness of the attack to 0\% while still maintaining the overall accuracy and only adding one additional call per question. The attack effectiveness for the \textit{CT with security agent} is shown in Table~\ref{tab:attack_summary}.

\begin{table}[h!]
  \centering
  \begin{tabular}{lccc}
    \toprule
    \textbf{Attack} & \textbf{CodeLlama} & \textbf{Mistral} & \textbf{GPT4.1-mini} \\
    \midrule
Baseline      & 73.17\(\pm4.01\) & 99.39\(\pm0.00\) & 99.19\(\pm0.87\) \\
Attack 1      & 63.21\(\pm7.16\) & 0.00\(\pm0.00\)  & 0.00\(\pm0.00\) \\
Attack 2      & -                 & -                 & 0.00\(\pm0.00\) \\
Attack 3      & 69.31\(\pm7.77\) & 0.00\(\pm0.00\)  & 2.44\(\pm1.51\) \\
Attack 4      & 88.01\(\pm4.87\) & 4.67\(\pm3.81\)  & 43.09\(\pm4.37\) \\
Attack 5      & 77.64\(\pm13.23\)& 3.05\(\pm1.51\)  & 15.85\(\pm6.60\) \\
Attack 6      & 48.37\(\pm9.74\) & 86.59\(\pm5.46\) & 71.95\(\pm1.51\) \\
    \bottomrule
  \end{tabular}
  \caption{\textit{Coder-tester with security analysis agent} tested on a baseline and six different attacks. The effectiveness (success rate) of each attack is shown. }
  \label{tab:attack_summary}
\end{table}


The baseline establishes the true negative rate (i.e., the agent correctly identified that no attack is present), which is 99.19\% for GPT-4.1-mini. Mistral performs similarly to GPT-4.1-mini at 99.39\%, while Code Llama is behind at 73.17\%. 
The true negative rate is acceptable for GPT-4.1-mini and Mistral.

Our best results are when we exploit the natural language understanding of the model, proving more successful than simply obfuscating or manipulating the code. 

The rows labeled \textbf{Attack} in Table~\ref{tab:attack_summary} determine the true positive rate, meaning the \textit{security analysis agent} correctly marked the code containing the injected function as malicious.
We found that GPT-4.1-mini performed well at detecting malicious code injections. Obfuscating the code, separating malicious code, dynamically changing the injected code, and others were all caught by GPT-4.1-mini. For example, \textbf{Attack 2}, which obfuscated the \textit{exfiltration function}, resulted in a 0\% attack success rate. The open-source models struggled with the obfuscated code, believing a base64 string to be an image, leading to undefined behavior. Their results for \textbf{Attack 2} are omitted from the table.

In the row labeled \textbf{Attack 1}, the effectiveness of the \textit{exfiltration function} with an added request library import and function call is shown. This attack is directly comparable to the attacks used on the \textit{coder-reviewer-tester}, which had an attack effectiveness of 6.71\% for GPT-4.1-mini. The added \textit{security analysis agent} was able to bring that down to 0\%. Likewise, Mistral also had an attack effectiveness rate of 0\% for \textbf{Attack 1}, a decrease of 95.73\% compared to the \textit{coder-reviewer-tester}. Code Llama performed worse, with a 63.21\% attack effectiveness, but still better than the \textit{coder-reviewer-tester}. 

\textbf{Attack 3} involved a comment stating that the \textit{exfiltration function} is commented out code, when in fact it is not. While the \textit{security analysis agent} believed that the uncommented code was actually commented out, it still decided to disapprove 97.56\% of the time, resulting in an attack success rate of 2.44\% for GPT-4.1-mini. Mistral outperformed GPT-4.1-mini, with an attack success rate of 0\%. However, in \textbf{Attack 4}, if the import statement for the requests library also has a comment above it claiming to be commented, the attack effectiveness increases greatly to 43.09\% for GPT-4.1-mini, but only 4.67\% for Mistral. Code Llama struggles with both attacks.

\textbf{Attack 5} uses a comment stating the \textit{exfiltration function} will not be executed and therefore should not be considered for review. This approach had limited success, 15.85\% for GPT-4.1-mini and 3.05\% for Mistral. Code Llama performed the worst at 77.64\%.

\textbf{Attack 6} injected a comment providing the model with poisonous few-shot examples. We used 9 few-shot examples, with 2 of them being correct and 7 of them being poisoned. We found that the correct examples were necessary for the attack to be successful. This attack achieved a success rate of 71.95\% for GPT-4.1-mini, while Mistral performed even worse at 86.59\%. Surprisingly, Code Llama was most resilient with an attack effectiveness of 48.37\%. 

Mistral outperformed or tied GPT-4.1-mini on the baseline and 5 of the 6 attacks. Code Llama performed the worst, likely due to it not being able to follow directions as well as the other models. While this negatively impacts its results, it may have helped on \textbf{Attack 6}, where it achieved the lowest attack effectiveness. The attack results show that while the \textit{security analysis agent} is more effective at preventing attacks than the \textit{coder}, \textit{coder-tester}, and \textit{coder-tester-reviewer}, it still is insecure if code is injected strategically.

\section{Limitations} \label{Limitations}

Our experiments were conducted using Code Llama 7b and Mistral 7b, along with GPT-4.1-mini, due to resource and time considerations. Therefore, larger LLM models may have higher accuracy and be better able to detect malicious code. In addition, we only evaluated our data using one dataset, HumanEval. Answers are contained within one function, which is not a realistic industry coding practice. Generally, projects contain many files and thousands of lines of code, making the task of the reviewers and security agents much more difficult. With more code and files, it is much easier to inject a function. Another limitation was the number of trials. Due to time and resource constraints, we were only able to run 3 trials for each of our results. This resulted in a large margin of error. Another limitation is the prompting of the LLMs; it is plausible that our prompts could be improved or adjusted, leading to better results.

\section{Future Directions} \label{Future_Directions}
This section presents future directions for the project.

\begin{itemize}
      \item Exploring the results of code injection on more realistic coding projects. 
      \item Rewording the prompts of the security agent or the reviewer may lead to better performance.
      \item Researching new methods for tricking the reviewer into being deceived by the injected code. It would also be interesting to explore whether the comments could be integrated directly into the code, through variable names or string values.
    \end{itemize}

\section{Conclusions} \label{Conclusion}

Agentic AI and Multi-Agent Systems are increasingly being used for code generation. While such systems can generate code quite accurately, they are vulnerable to attacks, including code injection. This paper analyzed the vulnerability of multi-agent systems and found that the \textit{coder-reviewer-tester} architecture is more resilient to code injection attacks than both the \textit{coder} and \textit{coder-tester} architectures, but is less efficient at writing code. Following industry standards, we added a specialized security analysis agent, which mitigates the loss in efficiency and achieves lower attack success rates compared to the \textit{coder-reviewer-tester} architecture. Despite this, we demonstrate that even with the security analysis agent, this approach is still vulnerable to multiple types of code injection attacks. Our experimental evaluation suggests that additional approaches are needed to mitigate the effectiveness of attacks for MAS to be deployed without supervision.

\bibliographystyle{abbrv}
\bibliography{Ref}

@article{huang2024resilience,
  title={On the resilience of {LLM}-based multi-agent collaboration with faulty agents},
  author={Huang, Jen-tse and Zhou, Jiaxu and Jin, Tailin and Zhou, Xuhui and Chen, Zixi and Wang, Wenxuan and Yuan, Youliang and Lyu, Michael R and Sap, Maarten},
  journal={arXiv preprint arXiv:2408.00989},
  year={2024}
}

@article{he2025red-teaming,
  title={Red-teaming {LLM} multi-agent systems via communication attacks},
  author={He, Pengfei and Lin, Yupin and Dong, Shen and Xu, Han and Xing, Yue and Liu, Hui},
  journal={arXiv preprint arXiv:2502.14847},
  year={2025}
}

@article{zhou2025guardian,
  title={GUARDIAN: Safeguarding {LLM} Multi-Agent Collaborations with Temporal Graph Modeling},
  author={Zhou, Jialong and Wang, Lichao and Yang, Xiao},
  journal={arXiv preprint arXiv:2505.19234},
  year={2025}
}

@article{triedman2025multi-agent_systems,
  title={Multi-agent systems execute arbitrary malicious code},
  author={Triedman, Harold and Jha, Rishi and Shmatikov, Vitaly},
  journal={arXiv preprint arXiv:2503.12188},
  year={2025}
}

@article{huang2023agentcoder,
  title={Agentcoder: Multi-agent-based code generation with iterative testing and optimisation},
  author={Huang, Dong and Zhang, Jie M and Luck, Michael and Bu, Qingwen and Qing, Yuhao and Cui, Heming},
  journal={arXiv preprint arXiv:2312.13010},
  year={2023}
}

@article{jiang2024a_survey_code,
  title={A Survey on Large Language Models for Code Generation},
  author={JIANG, JUYONG and WANG, FAN and SHEN, JIASI and KIM, SUNGJU and KIM, SUNGHUN},
  journal={arXiv preprint arXiv:2406.00515},
  year={2024}
}

@misc{john2025owasp,
  author       = {{OWASP GenAI Security Project}},
  title        = {2025 Top 10 Risk \& Mitigations for {LLMs} and {GenAI} Apps},
  year         = {2025},
  howpublished = {\url{https://genai.owasp.org/llm-top-10/}},
  note         = {Accessed: 2025-09-16}
}

@article{narajala2025securing_ATFAA,
  title={Securing Agentic {AI}: A Comprehensive Threat Model and Mitigation Framework for Generative {AI} Agents},
  author={Narajala, Vineeth Sai and Narayan, Om},
  journal={arXiv preprint arXiv:2504.19956},
  year={2025}
}

@misc{MITRE_ATLAS,
  title = {MITRE ATLAS Matrix},
  author = {{MITRE Corporation}},
  year = {2025},
  howpublished = {\url{https://atlas.mitre.org/matrices/ATLAS}},
  note = {Accessed: 2025-06-26}
}

@inproceedings{wang2023a_review_on_code,
  title={A review on code generation with {LLMs}: Application and evaluation},
  author={Wang, Jianxun and Chen, Yixiang},
  booktitle={IEEE International Conference on Medical Artificial Intelligence (MedAI)},
  pages={284--289},
  year={2023},
  organization={IEEE}
}

@article{dong2024self-collaboration,
  title={Self-collaboration code generation via {ChatGPT}},
  author={Dong, Yihong and Jiang, Xue and Jin, Zhi and Li, Ge},
  journal={ACM Transactions on Software Engineering and Methodology},
  volume={33},
  number={7},
  pages={1--38},
  year={2024},
  publisher={ACM New York, NY}
}

@article{chen2021humaneval,
  title={Evaluating large language models trained on code},
  author={Chen, Mark and Tworek, Jerry and Jun, Heewoo and Yuan, Qiming and Pinto, Henrique Ponde De Oliveira and Kaplan, Jared and Edwards, Harri and Burda, Yuri and Joseph, Nicholas and Brockman, Greg and others},
  journal={arXiv preprint arXiv:2107.03374},
  year={2021}
}

@article{codellama,
  title={Code {Llama}: {Open} {Foundation} {Models} for {Code}},
  author={Roziere, Baptiste and Gehring, Jonas and Gloeckle, Fabian and Sootla, Sten and Gat, Itai and Tan, Xiaoqing Ellen and Adi, Yossi and Liu, Jingyu and Sauvestre, Romain and Remez, Tal and others},
  journal={arXiv preprint arXiv:2308.12950},
  year={2023}
}

@article{autogen,
  title={Autogen: Enabling Next-Gen {LLM} Applications via Multi-Agent Conversation},
  author={Wu, Qingyun and Bansal, Gagan and Zhang, Jieyu and Wu, Yiran and Li, Beibin and Zhu, Erkang and Jiang, Li and Zhang, Xiaoyun and Zhang, Shaokun and Liu, Jiale and others},
  journal={arXiv preprint arXiv:2308.08155},
  year={2023}
}

@misc{OpenAI_OpenSDK,
  title = {OpenAI Agents {SDK}},
  author = {{OpenAI}},
  year = {2025},
  howpublished = {\url{https://openai.github.io/openai-agents-python/}},
  note = {Accessed: 2025-09-16}
}

@article{murugesan2025rise,
  title={The rise of agentic {AI}: implications, concerns, and the path forward},
  author={Murugesan, San},
  journal={IEEE Intelligent Systems},
  volume={40},
  number={2},
  pages={8--14},
  year={2025},
  publisher={IEEE}
}

@misc{Gartner,
  title = {Top Strategic Technology Trends for 2025: Agentic {AI}},
  author = {Coshow, Tom and Gao, Arnold and Pingree, Lawrence and Verma, Anushree and Scheibenreif, Don and Khandabattu, Haritha and Olliffe, Gary},
  year = {2024},
  howpublished = {\url{https://www.gartner.com/doc/reprints?id=1-2K8Y7LEY&ct=250212&st=sb}},
  organization = {The Gartner Inc.},
  note = {Accessed: 2025-07-09 }
}

@ARTICLE{ISO_SDLC,
  author={},
  journal={ISO/IEC/IEEE 12207 First edition 2017-11}, 
  title={{ISO/IEC/IEEE} International Standard - Systems and software engineering -- Software life cycle processes}, 
  year={2017},
  volume={},
  number={},
  pages={1-157},
  doi={10.1109/IEEESTD.2017.8100771}}

@inproceedings{kumar2007securityDev_SLDC,
  title={Security in coding phase of {SDLC}},
  author={Kumar, R and Pandey, SK and Ahson, SI},
  booktitle={Third International Conference on Wireless Communication and Sensor Networks},
  pages={118--120},
  year={2007},
  organization={IEEE}
}

@ARTICLE{STRIDE,
  author={Torr, P.},
  journal={IEEE Security \& Privacy}, 
  title={Demystifying the threat modeling process}, 
  year={2005},
  volume={3},
  number={5},
  pages={66-70},
  doi={10.1109/MSP.2005.119}}

@article{shi2024code,
  title={From code to correctness: Closing the last mile of code generation with hierarchical debugging},
  author={Shi, Yuling and Wang, Songsong and Wan, Chengcheng and Gu, Xiaodong},
  journal={arXiv preprint arXiv:2410.01215},
  year={2024}
}

@inproceedings{almorsi2024guided,
  title={Guided code generation with {LLMs}: A multi-agent framework for complex code tasks},
  author={Almorsi, Amr and Ahmed, Mohanned and Gomaa, Walid},
  booktitle={12th International Japan-Africa Conference on Electronics, Communications, and Computations (JAC-ECC)},
  pages={215--218},
  year={2024},
  organization={IEEE}
}

@article{islam2024mapcoder,
  title={Mapcoder: Multi-agent code generation for competitive problem solving},
  author={Islam, Md Ashraful and Ali, Mohammed Eunus and Parvez, Md Rizwan},
  journal={arXiv preprint arXiv:2405.11403},
  year={2024}
}

@article{wang2024survey_autonomous,
	title = {A Survey On Large Language Model Based Autonomous Agents},
	volume = {18},
	number = {6},
	journal = {Frontiers of Computer Science},
	author = {Wang, Lei and Ma, Chen and Feng, Xueyang and Zhang, Zeyu and Yang, Hao and Zhang, Jingsen and Chen, Zhiyuan and Tang, Jiakai and Chen, Xu and Lin, Yankai and {others}},
	year = {2024},
	note = {Publisher: Springer},
	pages = {186345},
}

@misc{freecodingtools_python_obfuscator,
  author       = {{Free Coding Tools}},
  title        = {Python Obfuscator},
  howpublished = {\url{https://freecodingtools.org/tools/obfuscator/python}},
  note         = {Accessed: 2025-07-09},
  year         = {2025},
}

@inproceedings{hong2024metagpt,
  title={Meta{GPT}: Meta programming for a multi-agent collaborative framework},
  author={Hong, Sirui and Zhuge, Mingchen and Chen, Jonathan and Zheng, Xiawu and Cheng, Yuheng and Zhang, Ceyao and Wang, Jinlin and Wang, Zili and Yau, Steven Ka Shing and Lin, Zijuan and others},
  year={2024},
  booktitle={International Conference on Learning Representations, ICLR}
}

\end{document}